\documentclass{article}
\usepackage[utf8]{inputenc}
\usepackage[margin=1.5in,top=1.0in,bottom=1.0in]{geometry}
\usepackage[english]{babel}
\usepackage[autostyle,english=american]{csquotes}
\usepackage{graphicx}
\MakeOuterQuote{"}

\usepackage[super,sort&compress]{natbib}
\usepackage[convert-footnotes=true,use-sort-key=false]{notes2bib}
\bibnotesetup{
  note-name = ,
  use-sort-key = false
}
\begin{document}

{\centering
{\Large Evolutionary forces in language change}\par\bigskip\par
Christopher A. Ahern$^{1}$, Mitchell G. Newberry$^{2}$,
  Robin Clark$^{1}$, Joshua B. Plotkin$^{2}$\par\medskip\par
  {\footnotesize Departments of $^1$Linguistics and $^2$Biology, 
    University of Pennsylvania, 19104}
  \par\medskip\par August 2016
\par}

\renewcommand{\abstractname}{}
\begin{abstract}
\noindent Languages and genes are both transmitted from
generation to generation, with opportunity for differential
reproduction and survivorship of forms.  Here we apply a rigorous
inference framework, drawn from population genetics, to
distinguish between two broad mechanisms of language change:
drift and selection.  Drift is change that results from
stochasticity in transmission and it may occur in the absence of
any intrinsic difference between linguistic forms; whereas
selection is truly an evolutionary force arising from intrinsic
differences -- for example, when one form is preferred by members
of the population.  Using large corpora of parsed texts spanning
the 12th century to the 21st century, we analyze three examples
of grammatical changes in English: the regularization of
past-tense verbs, the rise of the periphrastic `do', and
syntactic variation in verbal negation.  We show that we can
reject stochastic drift in favor of a selective force driving
some of these language changes, but not others.  The strength of
drift depends on a word's frequency, and so drift provides an
alternative explanation for why some words are more prone to
change than others.  Our results suggest an important role for
stochasticity in language change, and they provide a null model
against which selective theories of language evolution must be
compared.
\end{abstract}

\bigskip

There is a rich history of exchange between linguistics and
evolutionary biology, dating to the works of Darwin and
Haeckel\cite{schleicher1853ersten, darwin1888descent,
haeckel1868naturliche}.  While the mechanisms underlying
organismal evolution have been explored extensively, the forces
responsible for language evolution remain unclear.  Quantitative
methods to infer evolutionary forces developed in population
genetics have not been widely applied in linguistics, despite the
recent availability of massive digital
corpora\cite{michel2011quantitative, lin2012syntactic,
davies2010corpus, davies2012expanding}.

Language change can be viewed as competition between linguistic
forms, whether they are sounds, morphemes, or syntactic
structures\cite{cavallisforza-feldman1980cultural,
lieberman2007quantifying,pagel2007frequency,labov2010,
kroch1989reflexes,christiansen2016creating}.  The field of
linguistics has largely assumed that any substantial change in
the frequencies of alternative forms is due to selective forces
acting in the language community.  Linguists have identified many
sources of selection that could drive language change, including
language internal forces, cognitive forces, and social
forces\citep{paul1890principles, bloomfield1933,
jespersen1922language, fitch2010evolution, hock1991principles,
hock2009language, hoenigswald1978, jakobson1995language,
labov1994, labov2001, labov2010, wang1969,
weinreich1968empirical}.   It is unclear, however, whether these
are indeed the causative forces responsible for the changes
observed in languages over centuries.  To infer an evolutionary
force, we must first consider whether the observed changes are
due to stochasticity in transmission alone -- that is, drift.
Unlike selective forces, which bias a language learner towards
adopting forms that are intrinsically easier to learn or more
effective for communication, drift arises purely by chance: the
learner chooses randomly among the sample of forms that she
happens to encounter.  Although drift is recognized as an
important null hypothesis in population
genetics\citep{wright1931evolution} and cultural
evolution\cite{bentley2004random,cavallisforza-feldman1980cultural},
it has not yet been systematically tested in the context of
language change\cite{blythe2012neutral}.

\begin{figure}[t!] 
\centering
\includegraphics[width=0.8\linewidth]{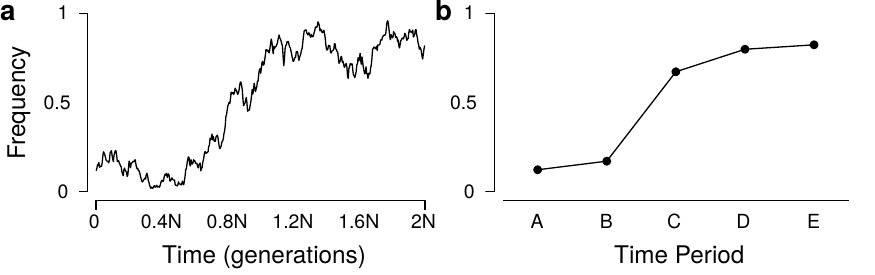}
\caption{\textbf{A null model of language change.}
Stochasticity in transmission can significantly change
the frequencies of alternative linguistic forms over
time, even without any intrinsic differences between
forms. We use the neutral Wright-Fisher diffusion from
population genetics\cite{crow-kimura1970}, which has
also been derived as a model of language
learning\cite{reali2010words}, as a null model of frequency
variation due to stochastic drift. Panel \textbf{a} illustrates
an example time-series of frequency variation produced by this
neutral null model.  Although the complete time-series evidently
shows random fluctuations, linguistic time-series require binning
texts by time period. When this time-series is binned (panel
\textbf{b}), it produces a characteristic $S$-shaped curve that
is often accepted as evidence of a selective force favoring one
linguistic variant over others\cite{baxter2009modeling,
blythe2012scurves, blythe2012neutral}. This simple example
illustrates the need to test hypotheses against a null model to
infer the presence of evolutionary forces in language
change\cite{blythe2012neutral}.}
\label{fig:drift}
\end{figure} 

\begin{figure}[t!] 
\centering
\includegraphics[width=0.8\linewidth]{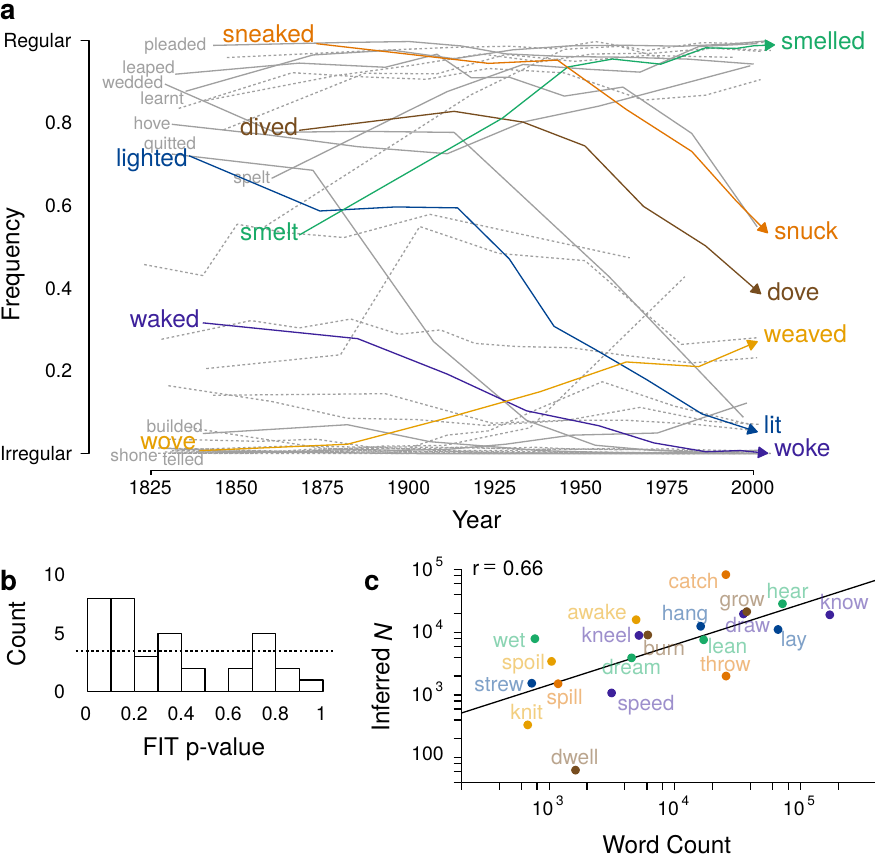}
\caption{\textbf{Verb regularization and irregularization.} We
analyzed 36 verbs with multiple past-tense forms appearing in the
Corpus of Historical American English\cite{davies2012expanding}.
Six of these verbs experience selection for either regularization
or irregularization, each with $p < $ 0.05 by the Frequency
Increment Test of selection\cite{feder2014identifying}
(\textbf{a}, colored lines).  The regular form is favored in two
of these cases, and the irregular form in the remaining four
cases.  Ten more verbs, of which four are regularizing
(\textbf{a}, solid gray lines), are significant at specificity
$1-\alpha = $ 0.8, with a false discovery rate of 45\%.  The
distribution of nominal FIT $p$-values (\textbf{b}) is
non-uniform (Kolmogorov-Smirnov $p = $ 0.002), which confirms
that some verbs experience selection.  Among the remaining 20
verbs, for which we fail to reject neutrality (\textbf{a}, dashed
gray lines), the log inferred population size correlates with log
token count in the corpora (\textbf{c}, Pearson $r=0.66,
p=0.002$).}
\label{fig:regularity}
\end{figure} 

\begin{figure}[t] 
\centering
\includegraphics[width=0.8\linewidth]{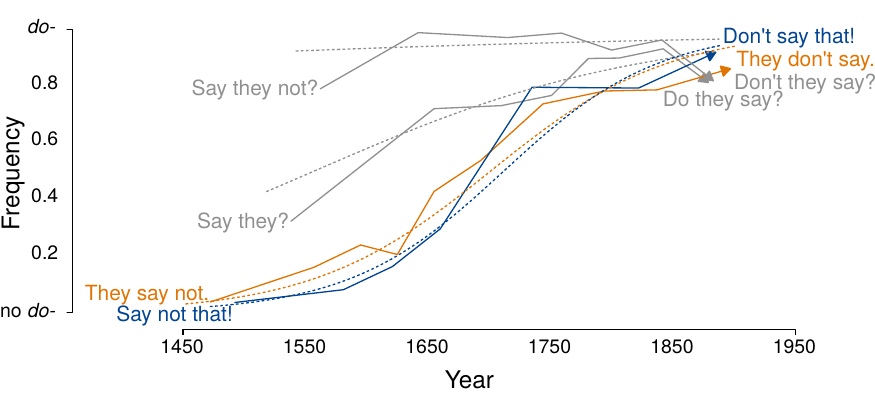}
\caption{\textbf{The rise of the periphrastic `do' in British
English}  The use of `do' as an auxiliary verb first arose in the
context of interrogative sentences (gray).  However, we cannot
reject drift for either affirmative interrogatives (4,401 cases
in parsed corpora, FIT $p=0.23$) or negative interrogatives (513
cases, FIT $p=0.77$).  Subsequently, the frequency of
\textit{do}-support rose rapidly in negative declarative (11,286
examples) and negative imperative (953 examples) sentences, where
we detect selection (FIT $p=0.01$ and $p=0.02$, respectively).
Dotted lines plot the logistic curve with slope determined by the
maximum-likelihood selection coefficient inferred in each
context.  Thus, \textit{do}-support may have arisen by chance in
interrogative statements, setting the stage for selection to
drive the evolution of \textit{do}-support in other grammatical
contexts.}
\label{fig:do}
\end{figure} 

\begin{figure}[t] 
\centering
\includegraphics[width=0.8\linewidth]{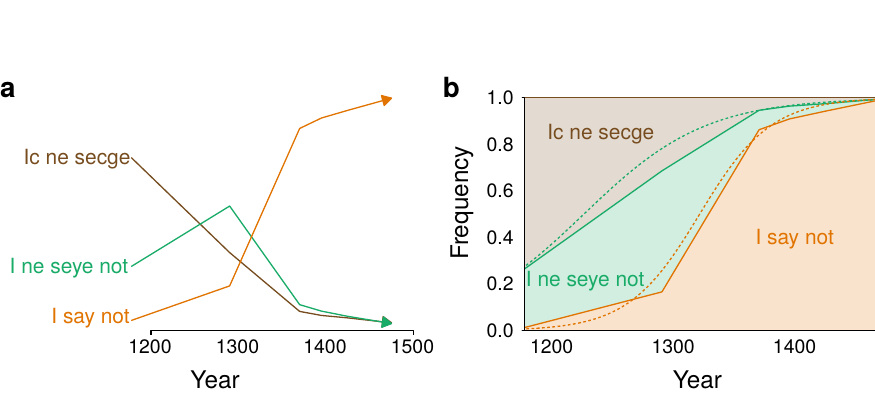}
\caption{\textbf{Evolution of sentential negation.} In English
and French, pre-verbal negation (e.g.~Old English "Ic ne secge")
gave way to embracing bipartite negation (Middle English "I ne
seye not") and then to post-verbal negation (Early Modern English
"I say not"), in a pattern known as Jespersen's Cycle.  We show
the frequencies these forms among 5,475 instances of negation
from 52 written works in the Penn-Helsinki Parsed Corpus of
Middle English (\textbf{a}).  We infer selection for bipartite
and post-verbal negation in the background of pre-verbal forms
(FIT $p = $ 0.03, \textbf{b}, green lines) and selection for
post-verbal negation in a mixed population of pre-verbal and
bipartite forms (FIT $p = $ 0.04, \textbf{b}, orange lines).
Dotted lines indicate logistic curves corresponding to
maximum-likelihood selection coefficients.}
\label{fig:jespersen}
\end{figure} 

Here we analyze three well-known grammatical changes in English:
the development of the morphological past tense in contemporary
American English\cite{pinker1991,cuskley2014} (spilt
$\rightarrow$ spilled); the rise of the periphrastic `do' in
Early Modern English\cite{ellegaard1953} (Mary ate not John’s
pizza $\rightarrow$ Mary did not eat John’s pizza); and
Jespersen's Cycle of sentential negation in Middle
English\cite{jespersen1917negation} (Ic ne secge $\rightarrow$ I
ne seye not $\rightarrow$ I say not).  Our analyses are based on
parsed English texts ranging from the Norman conquest of England,
during the 12th century, to the early 21st century. In each case,
we rigorously test whether observed linguistic changes are
consistent with neutral drift, or can be attributed to selective
forces.

We compare time-series of frequencies of alternative linguistic
variants to a null model of drift: the neutral Wright-Fisher
model from population genetics\citep{crow-kimura1970}.  The
Wright-Fisher model forms the basis for discriminating between
stochastic and selective forces on genetic variants in a
replicating population of size $N$; and the same model has been
derived for linguistic change under Bayesian
learning\cite{reali2010words}, where the inverse of the
population size parameter $N$ governs the amount of stochasticity
in transmission.  Importantly, even in the neutral case the
Wright-Fisher model can produce large frequency changes that may
appear, prima facie, to be the result of selection. It can even
produce the characteristic, logistic curve of one variant
replacing another in binned time-series (Figure~\ref{fig:drift})
that has typically been accepted as evidence of selective forces
in language change\cite{baxter2009modeling, blythe2012scurves,
blythe2012neutral}.

The population size parameter $N$ is unknown to us. And so to
infer the action of selection we must show that observed
linguistic changes are inconsistent with neutral drift,
regardless of $N$.  A stringent statistical test to reject this
composite null hypothesis ($s=0$, $N$ arbitrary) has recently
been developed, called the Frequency Increment Test
(FIT)\cite{feder2014identifying}.  The Frequency Increment Test
compares the frequency changes observed between sampled time
points to the expectations under drift. The test is valid for a
large class of neutral null models: all those with the same
diffusion limit as the Wright-Fisher model.  For each linguistic
time-series we can also estimate the most likely population size,
$N$, and the most likely selection coefficient, $s$, favoring one
linguistic variant over another\cite{feder2014identifying}.

We began by analyzing past-tense verb conjugation in contemporary
American English. One view contends that irregular past-tense
forms should regularize over time\cite{pinker1991, bybee2006,
bybee2001frequency}, for reasons of economy, phonological
analogy, or cognitive
ease\cite{jakobson1995language,zipf1949human,pinker1991}.  In
this view an irregular past-tense form, such as "wove," should be
selectively replaced by the regular form, "weaved", produced by
adding the voiced alveolar suffix "-ed" to the verb.  Although
there is substantial support for past-tense regularization,
especially for rare words over long
timescales\cite{michel2011quantitative,lieberman2007quantifying},
most studies have simply reported trends in usage frequencies
over time, and several apparent exceptions have been noted within
Modern English\cite{michel2011quantitative,cuskley2014}. 

We collected all past tense verb tokens from the Corpus of
Historical American English\cite{davies2012expanding},
comprising over four million words from $>100,000$ texts of
American English between the years 1810 and 2009, parsed for part
of speech.  Among all tokens assigned the simple past tense as
the most likely part of speech, we retained only those lemmas
with two variants each occurring at least 50 times in the
corpus\footnote{We also excluded lemmas with temporal variation
caused by spelling conventions (e.g~cancelled versus canceled),
lemmas with semantic ambiguities (e.g.~bear versus bore, wind),
and lemmas with multiple irregular variants (e.g.~begin, bid,
drink, ring).}. This produced 704,081 tokens in total which
provide frequency trajectories of regular versus irregular
past-tense variants for 36 verbs (Figure~\ref{fig:regularity})
\footnote{For each verb, we binned its tokens into date ranges of
variable lengths to ensure roughly the same number of tokens per
bin, setting the number of bins equal to the logarithm of the
number of tokens, rounded up. We applied Laplace (add-one)
smoothing to counts with only one of the two variants present, in
order to remove apparent absorption
events\cite{feder2014identifying}.}.

We used these linguistic time-series to determine whether
selection or drift is driving changes in past-tense conjugation
in Modern English (Figure~\ref{fig:regularity}). We computed a
two-sided $p$-value by the Frequency Increment Test for each of
the 36 verbs with irregular variants. For six of these verbs we
can reject neutral drift for all population sizes $N$, with
nominal $p<0.05$. Contrary to the standard linguistic
expectation, in four of these cases we infer selection towards
the \textit{irregular} variant (dived$\rightarrow$dove,
waked$\rightarrow$woke, lighted$\rightarrow$lit,
sneaked$\rightarrow$snuck), whereas only two cases exhibit
regularization (wove$\rightarrow$weaved,
smelt$\rightarrow$smelled).  Moreover, among the 16 verbs we
identify as possibly under selection, at specificity
$1-\alpha=0.8$ with a false discovery rate of 45\%, the majority
exhibit irregularization (Figure~\ref{fig:regularity}).  Examples
of irregularization have been noted previously, based on trends
in usage\cite{michel2011quantitative,cuskley2014}, whereas here
we have definitively inferred an evolutionary force operating on
Modern English verb conjugation (Kolmogorov-Smirnov $p$= 0.002,
Figure~\ref{fig:regularity}b)\footnote{Results of the Frequency
Increment Test for selection on these 36 verbs are not driven by
power or sample size. There is no significant difference in the
mean number of tokens among the 16 verbs with FIT $p<0.2$
compared to the remaining verbs (Mann-Whitney test, $p=0.29$).}.

Our analysis of irregular verbs illustrates the value of a null
model for language change.  Notably, for some verbs previously
described as undergoing
regularization\cite{michel2011quantitative, cuskley2014}, such as
spilt$\rightarrow$spilled and burnt$\rightarrow$burned, we cannot
reject neutral drift, even with sample sizes sufficient to reject
drift\footnote{The number of tokens for spill (1,178) exceeds
that of four verbs with FIT $p<0.2$. Likewise the number of
tokens for burn (6,097) exceeds that of eight verbs with FIT
$p<0.2$.}.  Conversely, we identify selection towards
irregularization in some cases, such as wedded$\rightarrow$wed,
that were previously predicted to be regularizing based on
long-term trends\citep{lieberman2007quantifying}.  We even
identify incipient grammatical changes, such as
wove$\rightarrow$weaved, in which the selected variant is in the
minority at present, but predicted by our analysis to eventually
replace the ancestral form.

Many studies have found that common words are more robust to
change over time than rare
words\cite{lieberman2007quantifying,bybee2001frequency,reali2009evolution,
fitch2007invisible}. Prevailing explanations for this phenomenon
are based on selection -- for example, purifying selection
against novel variants is assumed to be stronger for common
words\cite{pagel2007frequency}.  We propose an alternative and
complementary theory based on drift: more common words, whether
under selection to change or not, experience less stochasticity
in transmission.  This theory would predict less variability over
time among alternative variants of common words, even in the
absence of selection towards one form or another.  Indeed, we
find that for those past-tense verbs consistent with neutral
drift the most-likely inferred population size is correlated with
the word's frequency in the corpus (Pearson $r=0.66, p=0.002$,
Figure~\ref{fig:regularity}c).  Thus, the tendency for common
words to resist frequency
variation\cite{lieberman2007quantifying,bybee2001frequency,reali2009evolution,
fitch2007invisible} extends even to cases where we detect no
selective pressure for grammatical change. The relationship
between word frequency and the strength of drift also predicts
that different linguistic substitutions will occur by different
mechanisms: for rare words substitutions are more likely to occur
by random chance, whereas for common words substitutions are more
likely to be caused by selective forces.

Turning next to the rise of the periphrastic `do' in Early Modern
English, we collected tokens of \textit{do}-support from the
York-Helsinki Parsed Corpus of Early English Correspondence
(1400-1700), the Penn-Helsinki Parsed Corpus of Early Modern
English (1500-1700), and the Penn Parsed Corpus of Modern British
English (1710-1910), which include roughly seven million parsed
words from 1,220 texts of British English.  We extracted 16,072
tokens\cite{randall2009corpussearch2} of \textit{do}-support in
the context of affirmative questions, negative questions,
negative declaratives, and negative imperatives. Over the course
of these four centuries, for example, we see "You asked not."
become "You did not ask." and we see "Asked you a question?"
become "Did you ask a question?".

The rise of the periphrastic `do' in British English was more
rapid in negative declarative and imperative statements, where we
reject the neutral null model ($p<0.02$), than in interrogative
statements, where we fail to reject drift (Figure~\ref{fig:do}).
It may seem natural that selection for an auxiliary verb should
operate in all grammatical contexts
equally\citep{kroch1989reflexes}, and yet the extensive parsed
corpora available do not support this hypothesis. Our analysis
suggests an alternative scenario: the periphrastic `do' first
drifted by chance to high frequency in interrogative statements,
which then set the stage for subsequent selection in declarative
and imperative statements, for reasons of grammatical consistency
or cognitive ease. 

Finally, we studied the evolution of sentential negation from the
12th to the 16th century, based on 5,475 negative declaratives
extracted from the Penn Parsed Corpus of Middle English.  We
observe pre-verbal negation (e.g.~"Ic ne secge") giving way to
embracing bipartite negation ("I ne seye not") and then finally
to post-verbal negation ("I say not"), in a pattern known as
Jespersen’s Cycle\cite{jespersen1917negation}.  For both the
transitions that form this cycle we can definitively reject
neutral drift (FIT $p < 0.05$) in favor of a selective force
changing the formation of English negation
(Figure~\ref{fig:jespersen}). This quantitative analysis supports
longstanding linguistic hypotheses about forces driving verbal
negation, such as a tendency for speakers to use more emphatic
forms of negation\cite{ahern2015cycles, wallage2013functional,
dahl1979typology, dahl2001inflationary} which then becomes
normalized through frequent use by "pooling
action"\citep{ahern2015cycles, crawford-sobel1982strategic,
dahl1979typology, dahl2001inflationary,
kiparsky-condoravdi2006tracking, schwenter2006finetuning,
wallage2008jespersen, wallage2013functional}.

The field of comparative linguistics has long benefited from
quantitative techniques drawn from phylogenetics, producing a
detailed and nuanced characterization of the relationships
between different languages\citep{pagel2013linguistics,
fitch2010evolution, pagel2009human,pagel2007frequency}.  By
contrast, theories of how a given language changes over short
timescales have not been been subjected to quantitative inference
techniques. And yet, changes within a language must be the origin
of divergences and differentiation between languages.  Now, the
combination of massive digital corpora along with time-series
techniques from population genetics allows us to distinguish
hypotheses about the causes of language change from stochastic
drift, laying a foundation for empirically testable theories of
language evolution.

\pagebreak

\bibliographystyle{unsrtnat}

\begin{thebibliography}{57}
\bibitem[1()]{1}
We also excluded lemmas with temporal variation caused by spelling conventions
  (e.g~cancelled versus canceled), lemmas with semantic ambiguities (e.g.~bear
  versus bore, wind), and lemmas with multiple irregular variants (e.g.~begin,
  bid, drink, ring).
\bibitem[3()]{3}
Results of the Frequency Increment Test for selection on these 36 verbs are not
  driven by power or sample size. There is no significant difference in the
  mean number of tokens among the 16 verbs with FIT $p<0.2$ compared to the
  remaining verbs (Mann-Whitney test, $p=0.29$).
\bibitem[4()]{4}
The number of tokens for spill (1,178) exceeds that of four verbs with FIT
  $p<0.2$. Likewise the number of tokens for burn (6,097) exceeds that of eight
  verbs with FIT $p<0.2$.
\end{thebibliography}

\end{document}